\documentstyle[12pt]{article}

\def\beq{\begin{equation}}
\def\eeq{\end{equation}}
\def\beqa{\begin{eqnarray}}
\def\eeqa{\end{eqnarray}}

\def\bega{\begin{array}}
\def\enda{\end{array}}

\def\ll{{\langle}}
\def\rr{{\rangle}}
\def\lb{{\lbrack\!\lbrack}}
\def\rb{{\rbrack\!\rbrack}}
\def\pp{\rho}
\def\x{{\check x}}
\def\k{{\check k}}
\def\q{{\check q}}
\def\p{{\check p}}

\def\H{{\check H}}

\def\d{\partial}
\def\Schrodinger{{Schr\"odinger\ }}

\begin{document}

\title{Coupling ``Classical''\\ and Quantum Variables}
\author{Arlen Anderson\thanks{arley@physics.unc.edu}\\
Department of Physics and Astronomy\\
University of North Carolina\\
Chapel Hill NC 27510-3255 }
\date{Oct. 27, 1995}
\maketitle

\vspace{-10cm}
\hfill UNC-IFP 95-514

\hfill quant-ph/9511014
\vspace{10cm}

\begin{abstract}
Experimentally, certain degrees of freedom may appear classical because
their quantum fluctuations are smaller than the experimental error
associated with measuring them.  An approximation to a fully quantum
theory is described in which the self-interference of such
``quasiclassical'' variables is
neglected so that they behave classically when not coupled to other quantum
variables.  Coupling to quantum variables can lead to evolution
in which quasiclassical variables do not have definite values, but
values which are correlated to the state of the quantum variables.
A mathematical description implementing this backreaction of
the quantum variables on the quasiclassical variables is
critically discussed.
\end{abstract}
\newpage

%\baselineskip=24pt

%introduction

It is an observation of long standing that the world around us is (or appears
to be) largely classical. The fundamental description of the world is however
quantum mechanical. A natural and important question is whether one can
formulate an approximate description in which certain degrees of freedom
are treated as essentially classical while coupling them to other degrees
of freedom which are fully quantum. Such a description might be especially
important in exploring the domain between the fully classical and fully
quantum regimes. As well, it would be particularly useful in a subject like
quantum gravity where the full quantum theory is not known, and one cannot
make use of the semiclassical approximation. In both these cases, a problem
of particular interest is how one can describe and quantify the
backreaction of the quantum variables on the classical ones.  The positive
and negative features are discussed here of a proposal\cite{And} which gives
a mathematical prescription for coupling (quasi)classical
and quantum variables with physically desirable behavior.

The traditional approach to coupling classical and quantum variables is to
use expectation values wherever quantum variables appear in a
mixed set of equations of motion\cite{Kib}. This treats the full system
as essentially
classical and has the virtue of producing the realist-desired description
of a definite classical evolution. This approach can be criticized on a
number of grounds. In particular, an expectation value is not the outcome
of a single measurement but is an average of the outcomes of an ensemble of
identically prepared measurements. One might have expected that the
interaction with the classical variables was in some sense measuring the
quantum variables, but it is certainly not averaging over repeated
identical measurements\cite{repeat}. The result of this malapropos usage
of expectation value is that
this procedure gives physically wrong results when the expectation
value deviates from the most likely outcome(s) of a single measurement, as
it does for example in bimodal distributions.

An explicit example (cf. \cite{Kib}) illustrating the difficulty is given
by coupling the
momentum $p_a$ of a particle-a with the momentum $p_b$ of a second
particle-b through the interaction Hamiltonian $H_I=c p_a p_b$. Consider
first the fully quantum system, neglecting the self-Hamiltonians of
particle-a and -b. Suppose at time $t=0$ that the position of particle-a is
localized in a wavepacket $|\phi(x_a),0\rr$ with expected position $x_0$
and expected
momentum zero. Suppose also that at time $t=0$ particle-b is in a
superposition of two momentum eigenstates of equal and opposite momentum
$\bar p_b$
$$ |\psi,0\rr ={1\over 2^{1/2}}(|\bar p_b,0\rr + | -\bar p_b,0 \rr).$$
(This argument could be made with wavepackets for particle-b, but it is
easier to be explicit using eigenstates.)
A system initially prepared in a product state $|\phi,0\rr|\psi,0\rr$ will
evolve to a correlated superposition
\beq
\label{correl}
e^{-iH_It}|\phi,0\rr |\psi,0\rr=
 {1\over 2^{1/2}}(|\phi(x_a-c\bar p_b t),t\rr|\bar p_b,t \rr
     +|\phi(x_a+c\bar p_b t),t\rr|-\bar p_b,t \rr ).
\eeq
So, for example, if $\phi(x_a)=\pi^{-1/4}\exp(-(x_a-x_0)^2/2)$, then
$\phi(x_a-c\bar p_b t)=\pi^{-1/4}\exp(-(x_a-c\bar p_b -x_0)^2/2)$
is localized about $x_a=x_0 + c\bar p_b t$, as one would expect from the
solution of the Heisenberg equation of motion.

On the other hand, if particle-a were classical and one coupled its
position to the expectation value of the momentum of particle-b, there
would be no effect because
$$ \ll \psi,0 | p_b |\psi,0 \rr =0 .$$
This expectation value is the average of the two likely outcomes $\bar p_b$
and $-\bar p_b$ of a measurement.  It is not itself the outcome of any
measurement.  The classical particle is coupled to a phantom.  (The
situation would be more dramatic if the states were set up so that the
expectation value of $p_b$ in state $|\psi,0\rr$ were nonzero.)

A further
difficulty is exposed if one imagines that a momentum measurement is
subsequently made on $|\psi,t\rr$ and particle-b is projected into an
eigenstate
of definite momentum.  The expectation value of $p_b$ is suddenly nonzero and
the classical particle begins to feel the effect of the coupling.  This
is very peculiar behavior and would raise the relevance of the question
of when a measurement is completed to a daunting
level---it would have
physically meaningful consequences because the coupling
between classical and quantum systems would be changed by the act of
measurement.

These defects of coupling to expectation values are commonly interpreted as
evidence demonstrating the impossibility of coupling classical and quantum
variables.  This conclusion is too strong, but nevertheless the example
carries an important lesson about the nature of classical-quantum
interactions.  Consider what would happen if
particle-a were made increasingly classical starting from the fully
quantum result.  The state $|\phi,0\rr$ would go over into a
``state''$|(x_0,0),0\rr$
with position $x_a=x_0$ and momentum $k_a=0$.  The result of evolution
following from the classical limit of (\ref{correl}) is
\beq
 {1\over 2^{1/2}}(|(x_0+ c\bar p_b t, 0),t\rr|\bar p_b,t \rr
     +|(x_0- c\bar p_b t, 0),t\rr |-\bar p_b,t \rr).
\eeq
This has a ``classical'' particle in correlation
with the state of a quantum subsystem.  The ``classical'' particle-a
does not have a definite position.  Its specific location depends
on the quantum state. In this example that would not be determined
until the position of particle-a were observed or a momentum
measurement was made on particle-b.  Such  measurements
would show the position of particle-a to be correlated to the outcome of
the momentum measurement of particle-b as common sense would suggest.
An important and physically desirable feature of coupling classical
and quantum variables then is that it be possible for the value of a
``classical'' variable to depend on the
quantum state to which it is correlated.  Such a variable is not
classical in the realist sense of always having a definite value, so to
distinguish this, it shall be called {\it quasiclassical}.

One may well ask in what sense a variable is to be classical if it does
not take definite values.  The answer lies at the heart of the new
proposal.  A quasiclassical variable is one whose self-interference
effects can be neglected.  It is classical because it does not
exhibit observable interference phenomenon in its self-interaction.
When coupled to
a quantum system, the correlation with quantum states will generally
induce interference behavior on the quasiclassical variables, but it is
not an intrinsic property of those variables.  A mathematical encoding of
this definition will be proposed below, but it is valuable to elaborate
on its intuitive meaning first.

Every experiment has a scale of resolution or minimum experimental error
with which a measurement can be made. A quasiclassical variable is one
whose quantum fluctuations are negligible (or at least small) compared to
the experimental error with which the variable is known. This is
essentially an operational definition of what it means to appear classical.
No variable is actually classical; if examined closely enough, it will be
seen to have quantum fluctuations. But if the experimental error is
sufficiently large and the wavepacket not too delocalized, the quantum
fluctuations will essentially all take place within the error range where
they are indistinguishable from (classical) measurement uncertainty. In
that instance, the variable is operationally indistinguishable from being
classical. It is a stronger assumption that this condition persist under
evolution, but that is the property we desire of classical variables and
hence require of quasiclassical ones.

It should be emphasized that the apparent classical nature of a variable is
an experimental artifact. Consider the location of the center of mass of a
macromolecule of some extended size. The center of mass is not a
quasiclassical variable in and of itself simply because the mass is large.
Rather it is (if it is) because experiment fails to measure the location of
the center of mass to the necessary resolution to see quantum effects.
Arguably it is easier to measure the location of a concentrated point-like
object of a given mass than to measure the location of the center of mass
of a complicated extended object of the same mass. It may be that the
extended size and complex geometry of the macromolecule makes identifying
the location of the precise center of mass difficult. This is an
important remark because mathematically the center of mass variable behaves
like a point particle, but experimentally it is not observed as such.
Practically speaking, one is satisfied with knowing the macromolecule as a
whole is ``there,'' and the location of the molecule as seen in some
averaged sense is happily attributed to be that of the center of mass for
theoretical purposes. The motion of the molecule then behaves classically
because of the relatively imprecise limits that can be put on its position
and momentum.  Similar remarks would also hold for the other large scale
descriptors of the molecule like its linear dimensions, angular momenta,
etc.

The central argument that is exploited to understand the interaction of
quantum variables and quasiclassical ones is the following. Quasiclassical
variables, as actually part of a fully quantum system, are coupled to
other quantum
variables. This coupling can produce evolution which extends the wavepacket
of a quasiclassical variable beyond the range of its associated
experimental error. When this happens, the quasiclassical variable is in
correlation with the state of those other variables. If the coupling to the
other quantum variables were turned off, the quasiclassical variable would
be in a delocalized state which could be binned into a set of experimental
error intervals. Within each such interval the quasiclassical state would
be persistent by the assumption of negligible self-interference. It is thus
operationally classical within each interval. Which particular interval
occurs, or which set of intervals is possible, depends on the quantum state
to which the quasiclassical variable is correlated. As the knowledge of
this state is refined by measurement-observation, knowledge of the
quasiclassical variable is also refined.

One could preemptively observe the quasiclassical variable. Repeated
measurements of identically prepared situations would reveal that it does
not have the realist property of having a definite value (within
experimental error). This is expected: when correlated to other quantum
states, a quasiclassical variable need not be localized within a single
experimental error range.  Conventionally, one attributes this not to
the underlying quantum nature of
the quasiclassical variable, but to the correlated quantum states. These
states are viewed as the outcomes of a quantum ``event'' which triggered the
non-classical behavior.  The situation is the same as with Schr\"odinger's
cat.  From the fully quantum standpoint, this attribution is a
fiction, but in the quasiclassical framework it ``explains'' why more than
one outcome is possible for a classical object. Once the quasiclassical
variable is relocalized within a single measurement interval it will
persist within a neighborhood of that size until it is disrupted by
interaction with further quantum systems.

% quantum gravity

The paradigmatic example of a quantum event is that of a spin passing
through a Stern-Gerlach apparatus, and this will be discussed below. To
take a more extreme example to illustrate the significance of measurement
scales,  consider the case of gravity. Quantum gravitational
fluctuations are expected to be important at scales around the Planck length
($10^{-33}$~cm).
At length scales of general interest, they are many orders of magnitude
smaller than fluctuations of quantum matter variables.  Neglecting
quantum gravitational corrections to matter processes relative to
the contribution of quantum matter fluctuations is  generically
justifiable.  Since quantum gravitational
fluctuations are on a much smaller scale than can be seen experimentally,
and this
condition persists under ordinary evolution, one can ignore
the quantum nature of the gravitational field and treat the background
of spacetime as quasiclassical.

There is however the possibility of backreaction of the
quantum matter fields on the gravitational background. While quantum matter
fluctuations are very small on the length scales typically important for
classical gravity and their neglect is usually justified, these fluctuations
can lead
to qualitative changes in classical evolution, possibly by triggering
instabilities. This may be particularly important in the early universe. In
a different context, quantum fluctuations of a scalar field amplified by
inflation have already been proposed as the source of fluctuations in the
cosmic microwave background radiation and as seeds for galaxy
formation\cite{cmbr}.

% Choptuik example

A thought experiment makes the point sharper and again illustrates
the failing of the prescription of coupling to expectation values.
Choptuik\cite{Cho} has recently shown that classically a black hole forms from
spherically symmetric collapse of a massless scalar field whose initial
configuration is parametrized by a parameter $\pp$ when $\pp$ exceeds a
critical value $\pp^*$. For $\pp<\pp^*$, no black hole forms and the background
settles down to flat space as the scalar field disperses. Imagine a
wavepacket in $\pp$ of such initial configurations. Choose the wavepacket to
be localized so that it extends into the region above $\pp^*$ while
the expectation value of $\pp$ is less than $\pp^*$, $\ll \pp \rr < \pp^*$.
Coupling to the expectation value would lead to the
conclusion that no black hole forms.  Physical intuition leads one to
expect instead that a black
hole should form with a probability reflecting the likelihood of finding
the scalar field with $\pp>\pp^*$. One would say that quantum fluctuations of
the scalar field--reflected by the  nonvanishing amplitude of the
wavefunction above the critical value--lead to formation of the black hole.
Clearly, once a black hole forms, subsequent evolution in its presence will be
qualitatively different from evolution in flat space. It is to be able to
compute the probabilities of such events that a means of coupling
quasiclassical and quantum variables is needed.

% the formalism

The mathematical implementation of these ideas is comparatively simple at
first sight, while closer analysis reveals a number of subtleties. Consider for
convenience a system consisting of one quasiclassical degree of freedom and
one quantum degree of freedom. The extension to many variable systems is
straightforward. In brief, one has a pair of quantum canonical variables
$(\q,\p)$ satisfying the canonical commutation relation $[\q,\p]=i\hbar$
and a commutative pair of quasiclassical canonical variables $(\x,\k)$
satisfying a classical Poisson bracket relation $\{\x,\k\}=1$. Analogy to
the canonical commutation relations for a two-variable quantum system suggests
it is natural to assume all of the canonical variables commute except $\q$,
$\p$. This enables one to define functions of the canonical variables. The
Hamiltonian is such a function, $\H=H(\x,\k,\q,\p,t)$. If one forms the
coupled Heisenberg-Hamilton equations using this Hamiltonian, one has the
equations (at the initial time)
\beqa
\label{HH}
\dot q(t) |_{t=0}={-i\over \hbar}[\q,\H],&\quad&
\dot p(t) |_{t=0}={-i\over \hbar}[\p,\H], \\
\dot x(t) |_{t=0}= \{\x,\H\},&\quad& \dot k(t)|_{t=0}= \{\k,\H\}, \nonumber
\eeqa
where $q(0)=\q, p(0)=\p, x(0)=\x, k(0)=\k$.

The evolved variables $q(t), p(t), x(t)$, and $k(t)$ are in general
functions of $\q,\p,\x,\k$ and $t$.
While they divide into canonically conjugate pairs of purely quantum
and purely quasiclassical type at
the initial instant, once interaction begins, they
generally lose their particular identification as purely quantum
or quasiclassical, though they maintain their canonical
conjugacy.  This is a consequence of the coupling and is what
enables the quasiclassical variable to come into correlation with the
quantum state.  Note that there will always be some combination of the evolved
variables which form purely quantum and purely quasiclassical pairs, but
generally not $(q(t),p(t))$ and $(x(t),k(t))$.
This is the initial structure of the quasiclassical theory, and everything is
fairly straightforward. The subtleties begin to appear as one looks
closer.

% states and action of \x,\k as operators which read out eigenvalues

% uncertainty principle and relation to experimental measurement
%     uncertainty interval---in the quasi-classical variable one is looking
%     on such coarse scales that one doesn't see the quantum limit

First, the question of states must be addressed.  The quantum canonical
variables $(\q,\p)$ are operators which act on states in a
Hilbert space, as well as being algebraic elements with the canonical
commutation relations.  Some similar structure is needed for the
quasiclassical variables to act upon.  This has not been fully worked
out, but the natural starting point is to treat $\x$ and $\k$ as
acting on states $|(x',k'),0\rr$ as multiplication operators,
\beq
\x|(x',k'),0\rr =x'|(x',k'),0\rr,\quad \k|(x',k'),0\rr =k'|(x',k'),0\rr.
\eeq
Despite this ``operator'' nature of $\x$ and $\k$, for correspondence with
familiar experience, the term operator will be reserved to functions
involving the q-number operators $\q$ and $\p$ (which may involve $\x$ and
$\k$ as c-number parameters). The nature of the states associated with the
quasiclassical variables in the Schr\"odinger picture is unclear at the
present time, and, in case of confusion, it is recommended that one use the
Heisenberg picture where the states can be defined as ordinary joint
probability distributions in $(x',k')$ at the initial instant.

A key remark is necessary at this point about the uncertainty principle
with respect
to quasiclassical variables. The impression may be given by the notation
that the values of both $\x$ and $\k$ are known with infinite precision.
This is a false impression. As discussed above, in a real measurement
situation, there is an experimental resolution, or an experimental
error, to which variables are observed. The fact
that a variable has been identified as quasiclassical means that its
quantum fluctuations are persistently localized inside such an interval.
This in turn implies that one is well above the quantum limit when
observing that variable. The variable appears classical precisely because
one is not observing it too closely. In the quasiclassical approximation,
one idealizes the variable as fully classical (when not interacting with
quantum variables), but this is of course only a useful fiction. One cannot
turn around and attempt to measure the variable more closely, or the
quasiclassical approximation will break down. It is possible that it will
prove useful to implement a coarse-graining on the scale of the
experimental error to discourage attributing
significance to fine structure in the quasiclassical variable state on
scales smaller than this. The interplay between the experimental resolution
and the mathematical formalism representing the quasiclassical variables is
an aspect of this approach which needs further analysis.

Turn attention to the treatment of dynamics in this formalism. The
first point is that the Poisson bracket is defined as
\begin{equation}
\{ f,g \}= {\d f \over \d \x} {\d g \over \d\k} -{ \d f \over \d\k}
   {\d g \over \d\x}.
\end{equation}
By analogy to a two-variable classical system, it is assumed that the
$\x$ and $\k$ derivatives of $\q$ and $\p$ are zero.  This means that one
can compute, for example,
$$ \dot x(t)|_{t=0}={\d H \over \d\k}.$$
This will not be a c-number if a q-number multiplies a function of $\k$
in $H$.
The time derivative of a ``classical'' quantity needn't be a c-number!
This is precisely what enables the quasiclassical
variables to correlate with the state of the quantum ones.

A simple example will dramatize this.  Suppose that one couples a
spin-1/2 particle to a quasi-classical particle through the Hamiltonian
$H_I=c \k \sigma_z$.  The equations of motion (neglecting the self-Hamiltonian
for the quasiclassical particle) are
\beq
\dot x(t)= c \sigma_z,\quad \dot k(t)=0.
\eeq
The solutions to the equations of motion are
\beq
x(t)=\x + c\sigma_z t, \quad k(t)=\k.
\eeq
The solution for $x(t)$ involves the operator $\sigma_z$.

Suppose that the initial state of the system is given by the product state
\beq
|(x',0),0\rr |+x \rr,
\eeq
with the spin oriented in the $+x$ direction and the particle initially
at rest.  The operator nature of $x(t)$ can be interpreted by decomposing the
the quantum state into eigenfunctions of the operator component of $x(t)$.
The operator then returns a c-number eigenvalue
for each component, and a probability that that eigenvalue will be realized.
Here, one decomposes $|+x\rr$ into eigenstates of $\sigma_z$ and
finds the evolved state in the \Schrodinger picture to be
\beq
    {1\over 2^{1/2}}\biggl( |(x'+ct,k'),t\rr|\uparrow \rr
	+ |(x'-ct,k'),t\rr|\downarrow \rr\biggr).
\eeq
There is a probability of $1/2$ that the
quasiclassical particle will have either position $x'\pm c t$ at time $t$,
depending on the state of the spin to which it is correlated.

As discussed above, the quasiclassical variable will have an associated
experimental error.  The two possible outcomes for the position of the
quasiclassical particle will not be
distinguishable until their centers have separated by more than this
amount, and they can be resolved.  After they are capable of being
resolved, one has a superposition of quasiclassical (``macroscopic'') states
correlated to quantum states.  This is the same situation as with
Schr\"odinger's cat.  By observing either the quasiclassical state or the
spin, one destroys the superposition.  One interprets the multiple
possible quasiclassical outcomes as a consequence of the quantum
``event'' of the passage of the spin through the
magnetic field implicit in the interaction Hamiltonian.

The situation in the general case is similar to this. By decomposing the
quantum state into eigenfunctions of the operator part of the observable
of interest, one can determine the possible values that the observable takes
and with what probability. This is of course exactly the procedure one
takes to predict the possible outcomes of a measurement in a fully quantum
problem. If the quasiclassical state is initially in a joint probability
distribution and not specified by a specific value, then one must also take
this into account when determining the possible values of the
quasiclassical variables in the observable.

Return to the general issue of dynamics, and consider again the
equations of motion (\ref{HH}).  These are not sufficient in themselves
to determine the full evolution in general.  Suppose one wanted to
compute the second time derivative of $\x$ at $t=0$.  This should be given
by the
bracket of $\dot x$ with the Hamiltonian, but what bracket?  The first
derivatives were easy to compute because they each involved a canonical
variable of either purely quantum or purely quasiclassical type.  If
there is nontrivial coupling between the quasiclassical and quantum
variables, generally the first derivatives will be a mixture of
quasiclassical and quantum variables.  It is necessary to define the
bracket between two such mixed expressions.

Because quantum factor ordering information is lost in the classical
limit, as one canonical pair becomes quasiclassical, the quantum
canonical bracket does not have a unique correspondence to a
quasiclassical bracket.  This is the familiar problem in the
classical-quantum  correspondence.  There are two comparatively natural
candidates for quasiclassical brackets.

One is the
quasiclassical bracket proposed in \cite{And}.  For $A,B$ functions of the
quantum and quasiclassical variables,
\beq
\label{qcb}
\lb A,B \rb_{A}= {1\over i\hbar}[A,B] +
\biggl({\d A\over \d\x}{\d B\over \d\k} -
{\d A\over \d\k}{\d B\over \d\x}\biggr).
\eeq
If $A=Uf$ and $B=Vg$, where $U,V$ are functions of
$\q,\p$ and $f,g$ are functions of $\x,\k$, this takes the form
\beq
\label{qcb2}
\lb Uf,Vg \rb_{A}= {1\over i\hbar}[U,V] fg+i\hbar UV \{f,g\}.
\eeq
This bracket is not antisymmetric and hence not hermitian.

A second bracket, which is antisymmetric and hermitian, is the bracket
proposed independently by Alexandrov\cite{Ale} and by Boucher and
Traschen\cite{BoT} (ABT).  For $A,B$ functions of the
quantum and quasiclassical variables,
\beq
\label{ABT}
\lb A,B \rb= {1\over i\hbar}[A,B]
+{1\over 2}
\biggl({\d A\over \d\x}{\d B\over \d\k} -{\d A\over \d\k}{\d B\over \d\x}
+{\d B\over \d\k}{\d A\over \d\x} -{\d B\over \d\x}{\d A\over \d\k}\biggr).
\eeq
If $A=Uf$ and $B=Vg$, this is
\beq
\label{ABT2}
\lb Uf,Vg \rb= {1\over i\hbar} [U,V]fg +{1\over 2}(UV+VU) \{f,g\}.
\eeq

Both of these brackets give the correct relations among the canonical
variables $(\q,\p)$ and $(\x,\k)$, but note that the factor of $i\hbar$ has
been divided out of the purely quantum commutator. Both can be obtained by
taking the classical limit in an appropriate way\cite{And2}. Choosing one
imposes a canonical structure on the algebra of functions of all the
canonical variables.

An important issue is whether these brackets are derivations, that
is, whether they satisfy a product rule\cite{Dio},
\beq
\label{prod}
\lb A,BC\rb =\lb A, B\rb C + B\lb A,C\rb.
\eeq
The answer is that
neither is unconditionally a derivation\cite{And2}. The problem is that
taking the bracket of a variable is like taking a time derivative, and as
we have already seen, taking a time derivative can change a c-number into
something q-number valued. The result is that the factors in a product
which commute initially may not commute with the factors produced by taking
a derivative or bracket. Since the outcome depends on the order of factors,
a product rule will not hold in general.
A preferred ordering must hold initially to have a product rule.  By
choosing such an ordering, one is not affecting the value of the bracket,
only making it possible to evaluate with a product rule.

To be precise, consider the quasiclassical bracket (\ref{qcb}).  For
$A,B,C$ general functions of quantum and quasiclassical variables, one
finds\cite{And2}
\beq
\lb A, BC\rb_{A}= \lb A,B \rb_{A} C + B \lb A,C \rb_{A} + [ {\d
A\over \d\x},  B]  {\d C\over \d\k}  - [ {\d A\over \d\k} , B]
{\d C\over \d\x} .
\eeq
Since this bracket is not antisymmetric, there is a different rule when acting
from the right
\beq
\lb BC, A\rb_{A}= \lb B,A \rb_{A} C + B \lb C,A \rb_{A} +  {\d
B\over \d\x} [ C, {\d A\over \d\k} ] -  {\d B\over \d\k} [ C,
{\d A\over \d\x} ].
\eeq
If one decomposes $BC$ as a sum of terms of the form $fU$ with $f$ on the
left, where $U$ is quantum and $f$ is quasiclassical, then a product
rule holds in the first case.  In the second case, a product rule holds if
$BC$ is decomposed as a sum of terms $Uf$, with $f$ on the right.

A similar result holds for the ABT bracket (\ref{ABT}). There, one finds
\beqa
\lb A, BC\rb = -\lb BC,A \rb&=& \lb A,B \rb C + B \lb A,C \rb  + \\
&&\hspace{-3cm}{1\over 2} \biggl( [{\d A\over \d\x} , B] {\d C\over \d\k}  -
[{\d A\over \d\k},  B] {\d C\over \d\x}  +
{\d B\over \d\k} [C , {\d A\over \d\x} ]-
  {\d B\over \d\x} [C , {\d A\over \d\k}] \biggr).\nonumber
\eeqa
If one decomposes $BC$ as a sum of symmetrically ordered terms
${1\over 2}(fU +Uf)$, where $U$ is quantum and $f$ is quasiclassical,
then a product rule holds,
\beq
\label{prodABT}
\lb A, {1\over 2}(fU +Uf) \rb = {1\over 2} \biggl( \lb A,f \rb U +
f \lb A, U \rb + \lb A, U \rb f + U \lb A, f \rb \biggr).
\eeq

Because of the particular ordering of the quantum operators $U,V$ in
(\ref{qcb2}), the bracket is seen not to be antisymmetric and hence not
hermitian. This leads to the possibility that $\lb H,H
\rb_{A}\ne 0$, which in turn can lead to the peculiar situation that an
ostensibly time-independent Hamiltonian has a time-dependent evolution.
These features seriously complicate evolution and may be unphysical, so
this bracket will {\it
not} be used. The ABT quasiclassical bracket (\ref{ABT}) is antisymmetric
and hermitian and will be used for evolution.

Having chosen the quasiclassical bracket, one can now formulate the
derivative of a general time-dependent function. The equation of motion for
a function $A(q(t),p(t),x(t),k(t),t)$ with initial value $A(\q,\p,\x,\k,0)$
is
\beqa
{d A(q(t),p(t),x(t),k(t),t)\over dt}&=&
 \lb A(q(t),p(t),x(t),k(t),t),\H\rb  \\
&&\hspace{1.5cm}+ {\d A(q(t),p(t),x(t),k(t),t)\over \d t}, \nonumber
\eeqa
where $\H=H(\q,\p,\x,\k,t)$ is the Hamiltonian in terms of the initial
variables.  In particular, this gives the equations of motion for
$q(t), p(t), x(t),$ and $k(t)$,
\beqa
\label{HH2}
\dot q(t)=\lb q(t), \H \rb,&\quad&  \dot p(t)=\lb p(t), \H \rb, \\
\dot x(t)=\lb x(t), \H \rb,&\quad&  \dot k(t)=\lb k(t), \H \rb. \nonumber
\eeqa

It is very important to emphasize that $\H=H(\q,\p,\x,\k,t)$ is the
Hamiltonian expressed in terms of the initial variables. This is necessary
to be able to evaluate the bracket. If $A$ were expressed in terms of the
original variables, one could use (\ref{ABT}) to evaluate the bracket.
Alternatively, one could put $H$ into symmetrically ordered form and use
the product rule (\ref{prodABT}) to simplify the bracket. The ordering rule
which enables the bracket to satisfy a product rule is only known in terms
of the initial variables. This is because the multiplicative properties of
the canonical variables can change with time, so that one may have
$[x(t),k(t)]\ne 0$. The requirement that an expression be symmetrically
ordered as a product of a c-number and a q-number cannot be easily
satisfied in terms of the evolved variables.

Furthermore, $x(t)$ and $k(t)$ are not generally c-numbers, even if they
happen to mutually commute. One cannot take derivatives with respect to
them (without extending the definition of the derivative). This means
particularly that the quasiclassical bracket is not given in terms of the
evolved variables by an expression of the form (\ref{ABT}) with $\x,\k$
replaced by $x(t),k(t)$.

Nevertheless,  one
desires that the canonical relations between the canonical variables
computed with the quasiclassical bracket be preserved in time, e.g. $\lb
x(t), k(t) \rb =1$.  In purely quantum or classical theory, this follows from
the Jacobi identity for the bracket, but the Jacobi identity does not
hold in general for the quasiclassical bracket\cite{And4}.  One finds
\beqa
\lb \lb A, B \rb , C \rb - \lb \lb A, C \rb, B \rb - \lb A, \lb B, C \rb \rb
&=& \\
&&\hspace{-6.45cm}= \{ \{ A, B \}, C \} - \{ A, \{ B, C \} \}
- \{ \{ A, C \}, B \} + \{ A, \{ C, B \} \} \nonumber \\
&&\hspace{-6cm} \{ \{ B, C \}, A \} - \{ B, \{ C, A \} \}
- \{ \{ B, A \}, C \} + \{ B, \{ A, C \} \} \nonumber \\
&&\hspace{-6cm} \{ \{ C, A \}, B \} - \{ C, \{ A, B \} \}
- \{ \{ C, B \}, A \} + \{ C, \{ B, A \} \}. \nonumber
\eeqa
The right hand side of this equation would vanish if the Jacobi identity
were satisfied.  The main difficulty is the noncommutative nature of
$A,B,C$, but accepting the ordering it becomes as if one is missing part
of the Jacobi identity as it applies to the Poisson bracket. There are
obvious additional terms that one could add (maintaining ordering)
which would cause this to vanish, but there does not seem to be a
way to redefine the bracket so that they occur naturally.  For instance,
a term like $- \{ \{ A, C \}, B \}$ but where the differentiated
$B$ is ordered between $A$ and $C$ would cancel against the
first two terms on the right hand side.

For some Hamiltonians having special forms (particularly not coupling both the
coordinates and momenta of quasiclassical and quantum variables), a special
case of the Jacobi identity holds and it is sufficient to preserve
the brackets of the fundamental canonical variables.  One might conclude
that the quasiclassical approximation is not a good one for Hamiltonians not
of one of these forms.  Naturally one hopes that physically interesting
Hamiltonians are consistent, but this has not been proven and may not
be true.  Work is in progress to clarify this important issue.

The fact, $\lb H,H \rb=0$, implies that the only time-dependence $H$ has is
its explicit dependence. This is good because it means that
$$H(q(t),p(t),x(t),k(t),t)=H(\q,\p,\x,\k,t)=\H,$$
even though the detailed
expression of $H$ in terms of the evolved variables may have an ordering
which is not immediately obvious. The equation (\ref{HH2}) is not in fact
different from what one would naively expect.

The inconvenience of having to work with the initial variables is not as
serious as one might imagine. When solving the Heisenberg equations of
motion in quantum theory, one is trying to find the expression for the
evolved variables in terms of the initial ones. Having found a candidate
solution, the equations are verified by computing the commutator in the
initial variables. It is the same here.

Solutions to the equations of motion (\ref{HH2}) are most easily found by
developing a Taylor series expansion in time about the initial value. This
is done by evaluating higher time derivatives at the initial time by
taking further commutators with $H$. Since everything is evaluated at
the initial time, one can proceed iteratively with little difficulty
using (\ref{ABT})
to evaluate the bracket expressions. A second solution technique would be
to use canonical transformations\cite{And,And3}, but further work on this is
needed.

% conclusion?

The goal of the quasiclassical approach is to approximate a fully quantum
theory by treating approximately classical degrees of freedom as classical
when they are present in isolation yet coupling them to the quantum
variables in such a way that they may come into correlation with the
quantum state during interaction. The possibility of correlation between a
quasiclassical variable and the states in a quantum superposition is the
essential feature captured in this approach which is both observed
physically and yet is absent from the traditional semiclassical description
of coupling to the expectation value. The quasiclassical approximation is
implemented by neglecting the self-interference effects of degrees of
freedom which are persistently localized within their experimental
uncertainty.

A candidate mathematical approach to the quasiclassical approximation
treats the canonical conjugates associated to the
quasiclassical degrees of freedom as multiplicatively commutative
and retains
their canonical conjugacy through a classical Poisson bracket. This makes
these degrees of freedom behave classically in isolation. The coupling to
quantum degrees of freedom is accomplished by considering functions of both
commutative and noncommutative variables. A quasiclassical bracket is
defined which preserves the canonical structure of the classical and
quantum subalgebras and extends it to pairs of functions of the mixed set
of variables. This bracket is antisymmetric and hermitian and can be used
to define equations of motion which are essentially coupled
Hamilton-Heisenberg equations.  The complications are that the candidate
quasiclassical bracket satisfies a product rule only when acting on
quantities ordered in a particular way and the Jacobi identity does
not hold generally.   As a consequence, it is not certain how much of the
canonical structure is preserved under evolution.  The canonical relations
among the fundamental canonical variables are preserved for special
Hamiltonians, and work is in progress to determine for what class
of Hamiltonians this is true.

This work was supported in part by the National Science Foundation grant
PHYS 94-13207.


\begin{thebibliography}{99}
\bibitem{repeat}  The measurements may be thought of as being repeated in
time, but this is a separate kind of repetition already taken into account by
the changing of the
expectation value with time.  The repetition involved in computing the
expectation value refers to a collection of independent measurments made
on identically prepared systems at the same point in their evolution.
\end{thebibliography}
\end{document}